\documentclass{Interspeech2024}

\usepackage{multirow, amssymb}

\interspeechcameraready

\title{Zero-Shot Sing Voice Conversion: built upon clustering-based phoneme representations}

\name[affiliation={1}]{Wangjin}{Zhou}
\name[affiliation={2}]{Fengrun}{Zhang}
\name[affiliation={2}]{Yiming}{Liu}
\name[affiliation={3}]{Wenhao}{Guan}
\name[affiliation={2}]{Yi}{Zhao}
\name[affiliation={1}]{Tatsuya}{Kawahara}

\address{
  $^1$Kyoto University, Japan;
  $^2$Kwai Technology, China;
  $^3$Xiamen University, China;
  $^4$Tsinghua University, China
  }
\email{zhou@sap.ist.i.kyoto-u.ac.jp, 18811456416@163.com, liuyiming0053@163.com, whguan@stu.xmu.edu.cn, zhaoyi07@kuaishou.com, zy.2011@tsinghua.org.cn,kawahara@i.kyoto-u.ac.jp}


\begin{document}

\maketitle

\begin{abstract}
    This study presents an innovative Zero-Shot any-to-any Singing Voice Conversion (SVC) method, leveraging a novel clustering-based phoneme representation to effectively separate content, timbre, and singing style. This approach enables precise voice characteristic manipulation. We discovered that datasets with fewer recordings per artist are more susceptible to timbre leakage. Extensive testing on over 10,000 hours of singing and user feedback revealed our model significantly improves sound quality and timbre accuracy, aligning with our objectives and advancing voice conversion technology. Furthermore, this research advances zero-shot SVC and sets the stage for future work on discrete speech representation, emphasizing the preservation of rhyme.
\end{abstract}

\section{Introduction}

Singing Voice Conversion (SVC), a specific type of voice conversion, transforms a singing voice from one singer to another while retaining the original melody, lyrics, and musical nuances. The aim is to produce a new recording that convincingly simulates the target singer's performance without altering the musical composition. This paper explores a zero-shot any-to-any SVC framework noted for its robust generalization.

Singing Voice Conversion, similar to speaker voice conversion, focuses on audio reconstruction from content. However, given the complex rhythms and techniques in singing, standard Phonetic PosteriorGrams (PPG) inadequately capture essential content. Linguistic research highlights the International Phonetic Alphabet (IPA)'s limitations in capturing transitions between vowels and consonants, the effects of articulator interactions, and the inconsistent representation of phonetic features such as voicing and nasalization. Furthermore, the IPA inadequately reflects the nuanced phonological effects of articulation, limiting its effectiveness in detailing the intricate relationships among speech sounds~\cite{ladefoged1988some,browman1992articulatory,goldstein2003articulatory,mucke2020incongruencies}.

Content features from self-supervised learning (SSL) models~\cite{SSL2018} like WavLM~\cite{chen2022wavlm} and Hubert~\cite{hsu2021hubert} provide extensive information but also capture speaker-specific details. To ensure the reconstructed audio's timbre closely matches the target's, it's crucial to reduce timbral characteristics in content features. ContentVec~\cite{qian2022contentvec} utilizes a teacher-student approach with Hubert to separate speaker information from content, effectively reducing timbre leakage. FreeVC~\cite{li2023freevc} employs a spectrogram resizing technique and a bottleneck module with WavLM-extracted features, achieving good timbre replication in in-sample tests but less so in out-of-sample scenarios. KNN-VC~\cite{baas2023voice}, using clustered features from WavLM for the target speaker, significantly mitigates timbre leakage with minimal target speaker data within the RVC project\footnote{https://github.com/RVC-Project/}, though its performance on zero-shot conversion is hindered by its dependence on the amount of target data.

This research introduces a novel phoneme representation approach independent of existing linguistic frameworks, leveraging clustered features from multi-speaker data through the Hubert model for discrete, universal content representations. This method is grounded in two primary concepts: firstly, the Hubert model is a self-supervised learning (SSL) model optimized for extracting audio representations, highly beneficial for downstream tasks like automatic speech recognition, due to its ability to capture extensive content information. The impact of content loss through clustering is mitigated with an adequate number of cluster centers. Secondly, similar pronunciations across different speakers tend to converge within the same cluster centers, effectively abstracting pronunciation across speakers in a given frame while eliminating speaker-specific details.

Further, this study reveals that datasets with fewer recordings per singer are prone to timbre leakage, indicating that the strong in-domain voice conversion performance seen in prior research, which used open-source datasets with limited speaker variety but many recordings per individual, likely benefits from the extensive data on target speakers. Such studies, however, fall short in achieving effective out-of-domain, zero-shot voice conversion, indicating the need to address timbre leakage in creating versatile SVC models trained on large, diverse datasets.

Building on the so-vits-svc framework\footnote{https://github.com/svc-develop-team/so-vits-svc}, our approach utilizes clustering-based phoneme representation without relying on textual data, incorporating fundamental frequency values and speaker embeddings at the utterance level for singing voice reconstruction. Preliminary experiments on a dataset containing hundreds of hours of singing led to a refined model trained on an extensive compilation of 10,000 hours of audio from 500,000 singers. Demonstrating strong out-of-domain capabilities, our model meets the objective of zero-shot any-to-any singing voice conversion. Further details and demonstrations are accessible on our public demo page~\footnote{This paper demo page: https://yiathena.github.io/svc/}.
\begin{figure*}[t]
    \centering
	\includegraphics[scale=0.35]{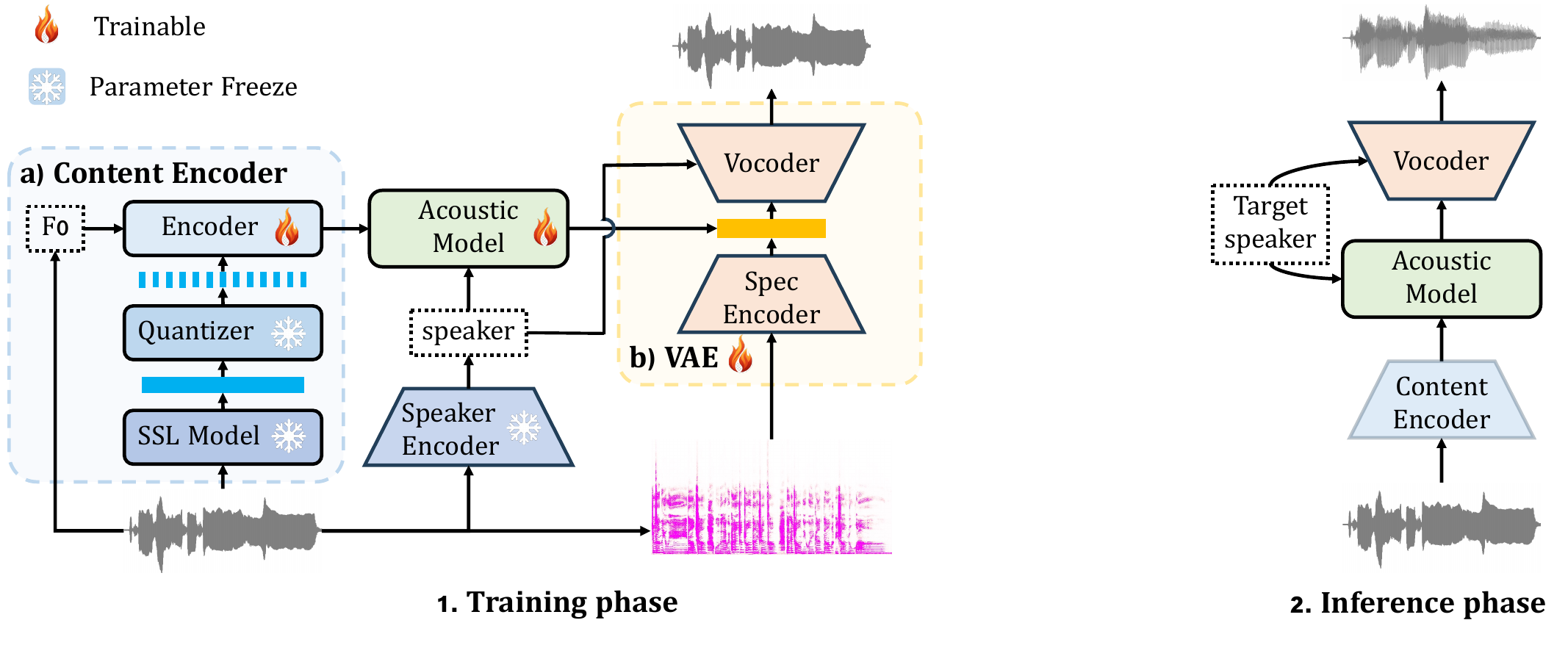}
	\caption{Basic Framework.}
	\label{fig:model_proposed}
 
\end{figure*}

\section{Proposed Approach}

As depicted in Figure~\ref{fig:model_proposed}, our research adopts a methodology consistent with the core architecture of non-autoregressive audio generation models. This approach includes a content encoder, an acoustic model, and a Variational Autoencoder (VAE)~\cite{mescheder2017adversarial}  focused on audio reconstruction. Furthermore, we incorporate WeSpeaker~\cite{wang2023wespeaker}  as a speaker encoder, significantly improving the model’s ability to identify and replicate speaker-specific features.

The training phase, described by Equation~\eqref{eq:train_all}, includes a two-step procedure: first, separating the original audio's components, and then, reconstructing the audio. In contrast, the voice conversion inference process, captured by Equation~\eqref{eq:infer_all}, focuses on producing audio in the target timbre. This involves extracting content information from the source audio and applying the target singer’s speaker embedding to this content.
  
\begin{align}
  Wav'_s = \mathbf{VC}(\mathbf{KM_{num}}(\mathbf{H}(Wav_s)), F0_s, S_s) \label{eq:train_all}\\
  Wav'_t = \mathbf{VC}(\mathbf{KM_{num}}(\mathbf{H}(Wav_s)), F0_t, S_t) \label{eq:infer_all}
\end{align}

Within this study, $\mathbf{H}$ represents the features extracted by the Hubert model, while $\mathbf{KM_{num}}$ denotes a k-means clustering algorithm with a specific number of cluster centers. The voice conversion (VC) module, $\mathbf{VC}$, illustrated in Figure \ref{fig:model_proposed}, integrates a trainable setup comprising an encoder, an acoustic model, and a Variational Autoencoder (VAE). The symbols $Wav_s$ and $Wav_t$ are used for the source and target waveforms, respectively. The output waveform is marked as $Wav'$. Pitch data from the source waveform is shown as $F0_s$, with $F0_t$ indicating the adapted pitch for the target waveform, adjusting $F0_s$ to align with the fundamental frequency of $Wav_t$. Lastly, speaker embeddings for the source and target are labeled as $S_s$ and $S_t$, respectively.

\subsection{Content Encoder}

The Content Encoder handles various content information, converting it into a format suited for the acoustic model. It utilizes the Hubert model to extract audio latent representation features, which are then discretized by a quantizer. Importantly, the self-supervised learning model, Hubert, along with the quantizer, are not involved in the training phase of the voice conversion module.

\subsubsection{Hubert}
Our study retrained the Hubert model on a vast, internally-curated dataset featuring 300,000 hours of multi-domain Chinese audio. We utilized the Hubert Large~\cite{hsu2021hubert} architecture, which includes seven CNN layers for downsampling and 24 transformer layers. Our initial research aimed to evaluate how different layers within the Hubert architecture affect audio reconstruction quality. We found that embeddings from the 22nd transformer layer yielded the most effective semantic representation for our objectives.

\subsubsection{Quantizer}
The function of the quantizer within the study is to discretize features extracted by the SSL model. We choose MiniBatchKMeans~\cite{chavan2015mini} as our quantization method for two primary reasons. Firstly, we posit that applying K-means clustering to a substantial multi-speaker dataset could identify cluster centers that encapsulate generalized phoneme information, effectively omitting speaker-specific nuances. Such discrete representations could then serve effectively as phoneme representations, especially since MiniBatchKMeans, when used with sufficiently large batch sizes, approximates the outcomes of traditional KMeans closely. Secondly, evidence from Base TTS~\cite{lajszczak2024base} indicates that vector quantization (VQ)~\cite{garbacea2019low} techniques reliant on back-propagation are significantly influenced by the distribution of data, which complicates the clustering of less prevalent phonetic features. However, our findings suggest that MiniBatchKMeans, particularly when trained with maximally large batch sizes, can mitigate the effects of an imbalanced data distribution.

To accommodate the processing of exceptionally large batch sizes swiftly, we develop a GPU-accelerated version of MiniBatchKMeans. This innovation enabled us to efficiently cluster data into 10,000 cluster centers with an unprecedented batch size of 1.5 million.

\subsubsection{Encoder}

The Encoder in this study is tasked with encoding a blend of diverse content information, with our experiments specifically focusing on the reconstruction of pitch information alongside the primary content. Specifically, the encoder is a multi-head attention.

\subsection{Acoustic Model}

The Acoustic Model is responsible for transforming content information into an intermediate representation for the VAE. 
We adopt a CNN-based flow structure~\cite{rezende2015variational}. It's important to note that the choice of acoustic model is versatile; options are not confined to this structure alone but can extend to encompass other types, including flow and diffusion models~\cite{ho2020denoising}, among others.
  
\subsection{VAE}

The VAE~\cite{mescheder2017adversarial} works as a waveform reconstructor, compressing the audio waveform into a compact, low-dimensional representation before reconstructing it back into its original waveform form. This compact representation serves as the crucial output for the Acoustic Model. We adopt WaveNet~\cite{van2016wavenet} as the spectral encoder and nsf-HiFiGan~\cite{wang2019neural} as the vocoder. It's important to stress that the choice of vocoder is not restricted to nsf-HiFiGan; a variety of vocoders, including various iterations of HiFiGan~\cite{lee2022bigvgan,kong2020hifi}, can be effectively utilized. 

\subsection{Speaker Encoder} 

The speaker encoder is designed to capture speaker-specific characteristics from singing voices. We choose ResNet34\footnote{https://wespeaker-1256283475.cos.ap-shanghai.myqcloud.com/models/voxceleb/voxceleb\_resnet34.zip}~\cite{he2016deep}, sourced from the WeSpeaker open-source framework, to serve as the speaker encoder. This encoded speaker information plays a crucial role in the functioning of both the acoustic model and the vocoder, ensuring that the unique vocal traits of each singer are accurately preserved and reflected in the synthesized audio.

\section{Experiments}
\subsection{Ablation Experiments}
We conduct multiple ablation studies: 

\noindent\textbf{a. uses only Hubert embeddings for reconstruction}
\begin{align}
  Wav'_s = \mathbf{VC}(\mathbf{H}(Wav_s)), \text{train} \label{eq:train1}\\
  Wav'_s = \mathbf{VC}(\mathbf{H}(Wav_s)), \text{infer} \label{eq:infer1}
\end{align}
In practical implementation, both f0 and speaker embeddings are set as tensors filled with ones.

\noindent\textbf{b. uses Hubert embeddings without clustering for SVC}
\begin{align}
  Wav'_s = \mathbf{VC}(\mathbf{H}(Wav_s), F0_s, S_s), \text{train} \label{eq:train2}\\
  Wav'_s = \mathbf{VC}(\mathbf{H}(Wav_s), F0_s, S_t), \text{infer1} \label{eq:infer21}\\
  Wav'_t = \mathbf{VC}(\mathbf{H}(Wav_s), F0_t, S_t), \text{infer2} \label{eq:infer22}
\end{align}

\noindent\textbf{c. uses clustered Hubert embeddings for SVC}
\begin{align}
  Wav'_s = \mathbf{VC}(\mathbf{KM_{num}}(\mathbf{H}(Wav_s)), F0_s, S_s), \text{train} \label{eq:train3}\\
  Wav'_s = \mathbf{VC}(\mathbf{KM_{num}}(\mathbf{H}(Wav_s)), F0_s, S_t), \text{infer1} \label{eq:infer31}\\
  Wav'_t = \mathbf{VC}(\mathbf{KM_{num}}(\mathbf{H}(Wav_s)), F0_t, S_t), \text{infer2} \label{eq:infer32}
\end{align}

Experiments \noindent\textbf{b.} and \noindent\textbf{c.} feature two inference modes based on whether the pitch (F0) has been modified through conversion. We assess the timbre retention by comparing the speaker cosine similarity between the generated waveform without pitch conversion $Wav'$ and the source waveform $Wav_s$ denoted as $SrcSIM$. This evaluation helps to determine the extent of timbre leakage. Similarly, the effectiveness of voice conversion is evaluated by comparing the speaker cosine similarity between the pitch-converted generated waveform $Wav'$ and the target waveform $Wav_t$, referred to as $TgtSIM$. Note that pitch conversion involves adding the difference between the mode of the F0 of $Wav_t$ and $Wav_s$ to $F0_s$, resulting in $F0_t$.

It is noted that since each experiment is conducted up to the 100,000th training step, the vocoder has not been fully optimized, resulting in the production of noisy singing voice outputs that could potentially influence MOS ratings during the ablation study phase. However, given that the training steps remained constant across all experiments, we consider the MOS scores to be comparable at this stage of the research.
  
\subsection{Dataset}

The data utilized for these ablation studies are segregated into two distinct datasets. Dataset A includes 50 hours of singing data acquired from an external source, featuring recordings from 100 singers, with each singer contributing an average of half an hour of samples. Conversely, Dataset B is comprised of 200 hours of singing data, internally collected from 10,000 non-professional singers, with each contributing approximately one minute of samples.

For the testing phase, the source samples are drawn from an internal collection distinct from Datasets A and B. This test dataset consists of 30-second clips from 80 different songs performed by 80 unique singers. The target speakers chosen for evaluating the test samples are selected for their diversity, including two male and two female singers, each represented by 10-second clips of their singing.

\subsection{Evluation Metrics}

We use Singer Similarity (SSIM) to assess the effectiveness of voice conversion by extracting speaker embeddings with a speaker verification (SV) model and computing the cosine similarity. We also use a 5-point (1-bad, 2-poor, 3-fair, 4-good, 5-excellent) Mean Opinion Score (MOS), which measured the naturalness. We selected 20 samples for these subjective evaluations and invited 25 evaluators to assess them.

\section{Discussions}

\subsection{Timbre Leakage}
\label{subsec:TimbreLeak}

As indicated in Table \ref{tab:results1}, Experiment~\textbf{a.} reveals that the Hubert features successfully capture a wealth of audio detail, including speaker identity information. Therefore, to ensure the similarity of timbre in the converted voice, it is crucial to disentangle speaker information from content features within representations from the Hubert model.

Experiment~\textbf{b.} shows that voice conversion models trained on datasets with limited recordings per singer lead to outcomes that resemble the original source waveform ($Wav_s$), regardless of the total data volume. This highlights the critical role of extensive per-speaker data in mitigating timbre leakage, particularly for models trained on publicly available datasets. Moreover, the findings indicate that the similarity to the target voice ($TgtSIM$) is less than the similarity to the source voice ($SrcSIM$) observed in Experiment ~\textbf{a.}, stressing the importance of addressing timbre leakage to enable zero-shot SVC with novel, unseen data.

A comparative analysis between Experiments \textbf{c.} and \textbf{b.} validates the effectiveness of the clustering approach introduced in this study in ameliorating the challenge of timbre leakage, marking a significant advancement in voice conversion task.

\begin{table}[t]
\setlength\tabcolsep{4.2pt}
\centering
\caption{Evaluations of timbre leakage. \textbf{AT} denotes ablation experiments type. H22 means embeddings extracted from the 22nd layer of Hubert. KM4096 menas Kmeans with 4096 cluster centers. Src SIM denotes speaker cosine similarity between $Wav'_s$ and $Wav_s$. Tgt SIM denotes speaker cosine similarity between $Wav'_t$ and $Wav_t$.
}
\small
\begin{tabular}{ccc|cc}
\hline
  \textbf{AT}  & \textbf{Content} & \textbf{Data} & \textbf{SrcSIM}$\downarrow$ & \textbf{TgtSIM}$\uparrow$ \\
\hline
a & H22 & B & 0.8149 & \textbackslash \\ \hline
\multirow{3}{*}{b} & H22 & A & $0.7350$ & $0.7819$ \\
                   & H22 & B & $0.7766$ & $0.7914$ \\ 
                   & ContentVec & B & $0.7073$ & $0.8216$ \\ \hline
c & H22+KM4096 & B & \textbf{0.6791} & \textbf{0.8517} \\ \hline
\end{tabular}
\label{tab:results1}
\end{table}
\vspace{0pt}

\subsection{Hubert Layer}
\label{subsec:HL}

The results in Table~\ref{tab:results2} arise from an initial exploration of the most effective layer within the Hubert model for extracting features, focusing on the comparative efficacy of layers 22 and 24 as shown in Figure 2.

Experiment~\textbf{a.} demonstrates that, although the 22nd and 24th layers contain comparable speaker information, the 22nd layer's embeddings are more conducive to reconstructing prosody. This advantage is primarily due to the lack of explicit pitch integration in the reconstruction process, which relies heavily on pitch cues within the Hubert embeddings for recreating singing. On the other hand, embeddings from the 24th layer provide less detailed prosodic information, resulting in reconstructions with a more mechanical sound.

Additionally, experiment~\textbf{c.} emphasizes the superiority of the 22nd layer in minimizing timbre leakage and improving speaker conversion accuracy. In contrast, reconstructions from the 24th layer are noted for their pronunciation of mechanical sense, as determined through subjective evaluation. Thus, this investigation firmly establishes the 22nd layer as the optimal choice for extracting content embeddings.

\begin{table}[t]
\setlength\tabcolsep{2.2pt}
\centering
\caption{Evaluations of Hubert layer comparison experiments. Subjective MOS with $95\%$ confidence intervals are shown. }
\small
\begin{tabular}{ccc|ccc}
\hline
  \textbf{AT}  & \textbf{Content} & \textbf{Data} & \textbf{SrcSIM}$\downarrow$ & \textbf{TgtSIM}$\uparrow$ & \textbf{MOS}$\uparrow$ \\
\hline
\multirow{2}{*}{a} & H22 & B & $0.8149$ & \textbackslash & $2.71\pm 0.09$\\
                   & H24 & B & $0.8168$ & \textbackslash & $2.44\pm 0.12$\\ \hline
\multirow{2}{*}{c} & H22+KM4096 & B & \textbf{0.6791} & \textbf{0.8517} & $3.83\pm 0.05$\\
                   & H24+KM4096 & B & $0.6903$ & $0.8475$ & $3.50\pm 0.07$\\ \hline
\end{tabular}

\label{tab:results2}
\end{table}

\subsection{Clustering Centers}
\label{subsec:CC}

To quantify clustering performance, three metrics are adopted for analysis: (1) AMD (Average Minimum Distance) which evaluates the clustering efficiency by measuring the average Euclidean distance from 1.5 million Hubert features to the nearest cluster center. (2) MDC (Minimum Distance among Cluster Centers) that assesses the granularity of phoneme distinction by the smallest distance between any two clusters. (3) QDC (Quintile Distance among Cluster Centers) that evaluates cluster separation resilience against outliers by the fifth percentile of the minimum Euclidean distances between cluster centers.

Findings in Table~\ref{tab:results3} show an inverse relationship between AMD and cluster numbers, indicating improved clustering precision with more clusters. Besides, the inverse trend between MDC and the number of clustering centers suggests enhanced detail resolution with more clusters.
At the same time, as cluster number increases, the relative change in QDC metrics is more stable compared to MDC, which reveals stable phoneme cluster structuring despite increased cluster numbers. 

In conclusion, the timbral replication remains stable across configurations, which supports that our model achieves great robustness in timbre preservation. In contrast, content fidelity improves with increased cluster numbers, enhancing the detail in rhythmic nuances. Notably, reducing clusters to 4096 compromises complex singing techniques and clarity in rapid speech, with a further reduction to 2048 exacerbating articulation issues.

\begin{table}[t]
\setlength\tabcolsep{2.2pt}
\centering
\caption{Evaluations of cluster centers.
}
\small
\begin{tabular}{ccc|ccc}
\hline \multicolumn{6}{c}{\small (a) VC Evaluation with 22nd Hubert Layer Embeddings}\\
\hline
  \textbf{AT}  & \textbf{Content} & \textbf{Data} & \textbf{SrcSIM}$\downarrow$ & \textbf{TgtSIM}$\uparrow$ & \textbf{MOS}$\uparrow$\\
\hline
\multirow{2}{*}{c} & KM10000 & B & $0.6923$ & $0.8500$ & $4.01\pm 0.04$ \\
                   & KM4096 & B & \textbf{0.6791} & $0.8517$ & $3.83\pm 0.05$ \\
                   & KM2048 & B & $0.6811$ & \textbf{0.8535} & $3.37\pm 0.07$ \\
\hline \multicolumn{6}{c}{\small (b) Cluster Evaluation }\\
\hline
  \textbf{AT}  & \textbf{Content} & \textbf{Data} & \textbf{AMD}$\downarrow$ & \textbf{MDC}$\downarrow$ & \textbf{QDC} \\
 & KM10000 & B & $307.95$ & $1073.00$ & $32876.69$ \\
 & KM4096 & B & $315.51$ & $3576.91$ & $39246.32$ \\
 & KM2048 & B & $326.83$ & $7418.21$ & $39503.28$ \\

\hline

\end{tabular}
\label{tab:results3}
\end{table}
\vspace{0pt}

\subsection{Final Model}
\label{subsec:FM}

For the final model, we compile a dataset from our internal archives, featuring 10,000 hours of singing from 500,000 individuals, with about a minute of data per person. We use a tenth of this dataset to train the clustering model with 10,000 centers via the MiniBatchKmeans method, with a batch size of 1.5 million. The training process is conducted over 10,000 iterations and takes about 48 hours on a single A100 GPU. The entire dataset is used to train the VC module. A Glow is adopted for the acoustic model, with nsf-HiFiGan as the vocoder. We modify the model parameters, setting $gin\_size$ to 256 and the vocoder's training slice length to 10,240, with a $hop\_length$ of 512. The VC module undergo training on 40 A100 GPUs with 80 mini batch size, requiring 350 hours for 800,000 steps.

Table \ref{tab:results_final} shows that our SVC model significantly distinguishes between source and target speakers in the test set, achieving high similarity with the target speaker and low similarity with the source. Content quality and reconstruction meet our standards. We also experiment with zero-filled tensors for speaker embedding, leading to a more neutral voice timbre as expected. However, this method results in reduced audio quality after conversion, due to the significant difference from the training sample embeddings.

\begin{table}[t]
\setlength\tabcolsep{4.2pt}
\centering
\caption{Final model performance. The empty speaker means inference with all zeros tensor as speaker embedding.}
\begin{tabular}{c|ccc}
\hline
\textbf{Speaker} & \textbf{SrcSIM}$\downarrow$ & \textbf{TgtSIM}$\uparrow$ & \textbf{MOS}$\uparrow$ \\
\hline
Tgt & $0.6658$ & $0.8614$ & $4.3\pm 0.05$\\ \hline

\end{tabular}
\label{tab:results_final}
\end{table}

\section{Conclusion}

We introduced a new Zero-Shot any-to-any SVC method that addresses limitations found in previous VC research using open-source datasets. While prior work demonstrated some out-of-domain VC capability, it benefited from having too many samples per speaker in the training dataset; we show that this capability diminishes with limited samples per speaker. Additionally, we investigated a clustering-based phoneme representation approach, examining its impact on audio reconstruction and clustering dynamics. Our work advances voice conversion technology by enhancing discrete speech representation and addressing key issues such as timbre leakage and phonetic representation, without heavily relying on pre-existing linguistic knowledge.

\newpage

\bibliographystyle{IEEEtran}
\bibliography{mybib}

\end{document}